\documentclass[12pt,a4paper]{article}
\usepackage{amsmath}
\usepackage{graphicx}
\usepackage{cite}

\allowdisplaybreaks[3]

\begin{document}

\begin{titlepage}

\begin{flushright}
SNUTP15-007, KUNS-2574
\end{flushright}

\vspace{5em}

\begin{center}
{\Large\bf Stochastic Dynamics of Infrared Fluctuations \\in Accelerating Universe}
\end{center}

\begin{center}
Gihyuk C{\sc ho}$^{1)}$\footnote{E-mail address: whrlsos@snu.ac.kr},  
Cook Hyun K{\sc im}$^{1)}$\footnote{E-mail address: protozerta@snu.ac.kr} 
and Hiroyuki K{\sc itamoto}$^{2)}$\footnote{E-mail address: kitamoto@tap.scphys.kyoto-u.ac.jp}
\end{center}

\begin{center}
$^{1)}${\it School of Physics and Astronomy, \\Seoul National University, \\Seoul 151-747, Korea}\\
$^{2)}${\it Division of Physics and Astronomy, \\Graduate School of Science, Kyoto University,\\Kyoto 606-8502, Japan}
\end{center}

\begin{abstract}
We extend investigations of infrared dynamics in accelerating universes. 
In the presence of massless and minimally coupled scalar fields, physical quantities may acquire growing time dependences through quantum fluctuations at super-horizon scales. 
From a semiclassical viewpoint, it was proposed that such infrared effects are described by a Langevin equation. 
In de Sitter space, the stochastic approach has been proved to be equivalent to resummation of the growing time dependences at the leading power. 
In this paper, we make the resummation derivation of the Langevin equation in a general accelerating universe. 
We first consider an accelerating universe whose slow-roll parameter is constant, and then extend the background as the slow-roll parameter becomes time dependent. 
The resulting Langevin equation contains a white noise term and the coefficient of each term is modified by the slow-roll parameter. 
Furthermore we find that the semiclassical description of the scalar fields leads to the same stochastic equation as far as we adopt an appropriate time coordinate. 
\end{abstract}

\vspace{\fill}

Sep. 2015

\end{titlepage}

\section{Introduction}\label{Introduction}
\setcounter{equation}{0}

In accelerating universes, the degrees of freedom at super-horizon scales increase with time. 
This increase gives rise to a growing time dependence to the propagator of a massless and minimally coupled scalar field. 
In some field theoretic models on the backgrounds, physical quantities acquire secular growths through the propagator. 
We call them quantum infrared (IR) effects in accelerating universes. 

For instance, in de Sitter (dS) space, the propagator of a massless and minimally coupled scalar field depends logarithmically on the scale factor of the Universe \cite{Vilenkin1982,Linde1982,Starobinsky1982}. 
By employing the Schwinger-Keldysh formalism \cite{Schwinger1961,Keldysh1964}, we can perturbatively investigate interacting field theories on a time dependent background. 
At each loop level, the IR effects in dS space manifest as polynomials in the logarithm of the scale factor. 
At late times, the leading IR effects come from the leading logarithms at each loop level. 
This fact indicates that perturbation theories eventually break down if the dS expansion continues long enough. 

From a semiclassical description of the scalar field, A. A. Starobinsky and J. Yokoyama proposed that the IR effects can be evaluated nonperturbatively by a Langevin equation \cite{Starobinsky1986,Starobinsky1994}.   
N. C. Tsamis and R. P. Woodard proved that the stochastic approach is equivalent to the resummation of the leading logarithms \cite{Woodard2005}. 
These investigations were performed in interacting field theories with potentials in dS space. 

In a general accelerating universe, it holds true that the propagator of a massless and minimally coupled scalar field is growing with time.  
Therefore, in a similar way to dS space, some nonperturbative analyses of the IR effects are necessary in interacting field theories with the scalar field. 

In this paper, we extend the resummation formula of the leading IR effects in a general accelerating universe. 
Although there have been some attempts to derive the stochastic equation in a general accelerating universe \cite{Starobinsky2009,Starobinsky2010-1,Starobinsky2010-2,Starobinsky2015}, 
the consistency with the resummation formula has not been verified completely.  
We confirm that in a general accelerating universe, the Yang-Feldman formalism is reduced to a Langevin equation with a white noise term up to the leading IR effects. 
Furthermore the resulting Langevin equation shows that in the semiclassical description of the scalar field, the choice of time coordinate is crucial to obtain the correct coefficient of its nonlinear term\footnote{
The dependence of the choice of time coordinate is discussed for the stochastic dynamics of inflatons \cite{Starobinsky2015}. 
Our claim is that it is crucial even for the stochastic dynamics of test scalar fields. }. 

As a first step, we consider an accelerating universe whose slow-roll parameter is constant. 
On the background, the wave function of a massless scalar field is expressed by the Bessel function.  
As utilized in \cite{Prokopec2008,Prokopec2009,Prokopec2015-1,Prokopec2015-2}, this fact facilitates analytic investigations. 
In Section \ref{FC}, we review the free scalar field theory on the background, especially its IR dynamics. 
In Section \ref{NC}, we first derive the Langevin equation in the accelerating universe by the resummation of the leading IR effects, 
and then confirm that the semiclassical description of the scalar field leads to the same equation if we choose an appropriate time coordinate. 

In the latter part of this paper, we remove the constraint of the slow-roll parameter. 
In a general homogeneous, isotropic and spatially flat space, the wave function of a massless and minimally coupled scalar field has been derived \cite{Woodard2003}.
In Section \ref{FG}, by reference to the previous study, we confirm that the propagator has a growing time dependence in a general accelerating universe.   
In Section {\ref{NG}}, we extend the nonperturbative analyses in Section \ref{NC} in a general accelerating universe.   
We conclude with discussions in Section \ref{Conclusion}. 

\section{Free scalar field theory in an accelerating universe with $\dot{\epsilon}=0$}\label{FC}
\setcounter{equation}{0}

In this paper, we consider a homogeneous, isotropic and spatially flat universe: 
\begin{align}
ds^2&=-dt^2+a^2(t)d{\bf x}^2 \notag\\
&=a^2(\eta)(-d\eta^2+d{\bf x}^2),  
\label{metric}\end{align}
where $t$ is a cosmic time and $\eta$ is a conformal time. 
The dimension of the spacetime is taken as $D=4$. 
We consider an expanding era when the Hubble parameter is positive 
\begin{align}
H\equiv\frac{\dot{a}}{a}>0. 
\end{align}
In our notation, the dot denotes the derivative with respect to $t$. 
In order to describe deformations from dS space, we introduce the slow-roll parameter whose parameter region is given by 
\begin{align}
\epsilon\equiv\frac{-\dot{H}}{H^2},\hspace{1em}0\le \epsilon<1. 
\label{acceleration}\end{align}
Here the lower limit corresponds to dS space and the upper bound means an accelerating universe. 

In Section \ref{FC} and \ref{NC}, we set the slow-roll parameter as it does not change in time
\begin{align}
\epsilon=\epsilon_0: \text{const.} 
\end{align}
On the background, the scale factor is written by the cosmic time and the conformal time as 
\begin{align}
a=(1+\epsilon_0H_0t)^\frac{1}{\epsilon_0}=\big\{\frac{1}{-(1-\epsilon_0)H_0\eta}\big\}^\frac{1}{1-\epsilon_0}. 
\label{a1}\end{align}
In the parameter region (\ref{acceleration}), the conformal time approaches zero at the infinite future. 
The Hubble parameter and the Ricci scalar are given by 
\begin{align}
H=H_0a^{-\epsilon_0},\hspace{1em}R=6(2-\epsilon)H^2. 
\end{align}
Here $H_0$ denotes an initial value of the Hubble parameter. 

Under the scale transformation: 
\begin{align}
\eta\to C\eta,\hspace{1em}{\bf x}\to C{\bf x}, 
\end{align}
the Hubble parameter and the metric of the background (\ref{a1}) scale as follows 
\begin{align}
H\to C^\frac{\epsilon_0}{1-\epsilon_0}H,\hspace{1em}ds^2\to C^{-\frac{2\epsilon_0}{1-\epsilon_0}}ds^2. 
\end{align}
The scaling is understandable since the dimension of each quantity is given as $[H]=1$, $[ds^2]=-2$. 
At quantum level, if a vacuum expectation value (vev) of a $D$-dimensional operator $[\mathcal{O}]=D$ is written as 
\begin{align}
\langle \mathcal{O}(x)\rangle\propto H^D, 
\label{scalinglaw}\end{align}
we call that the scaling law is respected. 
Actually the IR effects which we discuss in Section \ref{FC} and \ref{NC} can be identified as breakdown of the scaling law. 

We consider a massless and minimally coupled scalar field, the quadratic action of which is given by 
\begin{align}
S_2=\int \sqrt{-g}d^4x\ \big[-\frac{1}{2}g^{\mu\nu}\partial_\mu\varphi\partial_\nu\varphi\big]. 
\end{align}
For the free scalar field $\varphi_0$, we can expand it by the annihilation and the creation operators $a_{\bf p}$, $a_{\bf p}^\dagger$ as  
\begin{align}
\varphi_0(x)=\int\frac{d^3p}{(2\pi)^3}\ \big[a_{\bf p}\phi_{\bf p}(x)+a^\dagger_{\bf p}\phi^*_{\bf p}(x)\big],
\end{align}
\begin{align}
\phi_{\bf p}(x)&=a^{-1}\times\frac{\sqrt{\pi}}{2}(-\eta)^\frac{1}{2}H_{\nu_0}^{(1)}(-p\eta)e^{+i{\bf p}\cdot{\bf x}}, 
\end{align}
\begin{align}
\nu_0=\frac{3-\epsilon_0}{2(1-\epsilon_0)}=\frac{3}{2}+\frac{\epsilon_0}{1-\epsilon_0}, 
\label{nu1}\end{align}
where $H_{\nu_0}^{(1)}$ is the Hankel function of the first kind. 
The subscript $0$ of $\nu_0$ is inserted to emphasize its time independence. 
We consider the Bunch-Davies vacuum $a_{\bf p}|0\rangle=0$ in this paper. 
It should be noted that $p$ denotes the comoving momentum and the physical momentum is given by $P= p/a$.

At sub-horizon scales: $P\gg H\Leftrightarrow -p\eta\gg \frac{1}{1-\epsilon_0}$, 
the wave function approaches that in Minkowski space up to the conformal factor 
\begin{align}
\phi_{\bf p}(x)\sim a^{-1}\times\frac{1}{\sqrt{2p}}e^{-ip\eta+i{\bf p}\cdot{\bf x}-i\frac{2\nu_0+1}{4}\pi}. 
\end{align}
In this paper, we focus on contributions from super-horizon scales rather than those from sub-horizon scales. 
At the IR scales, specific behaviors for curved spacetimes become dominant. 

In investigating the IR behavior at super-horizon scales: $-p\eta\ll \frac{1}{1-\epsilon_0}$, 
the following expansion of the Bessel function is useful 
\begin{align}
H_{\nu_0}^{(1)}(z)&=J_{\nu_0}(z)+iN_{\nu_0}(z), \notag\\
N_{\nu_0}(z)&=\frac{1}{\sin(\pi\nu_0)}\big\{\cos(\pi\nu_0) J_{\nu_0}(z)-J_{-\nu_0}(z)\big\},
\label{ex1}\end{align}
\begin{align}
J_{\pm\nu_0}(z)=\big(\frac{z}{2}\big)^{\pm\nu_0}\sum^\infty_{n=0}\frac{(-1)^n}{n!\Gamma(1\pm\nu_0+n)}\big(\frac{z}{2}\big)^{2n}, 
\label{ex2}\end{align}
where the double-sign corresponds. 
The leading IR term of the wave function comes from the zeroth order of the expansion of $J_{-\nu_0}$: 
\begin{align}
\phi_{\bf p}(x)\simeq a^{-1}\times -i(-\eta)^\frac{1}{2}\frac{2^{\nu_0-1}\Gamma(\nu_0)}{\sqrt{\pi}}\frac{1}{(-p\eta)^{\nu_0}}e^{+i{\bf p}\cdot{\bf x}}. 
\label{w1}\end{align}

From (\ref{w1}), the contribution from super-horizon scales to the propagator is given by 
\begin{align}
\langle\varphi_0^2(x)\rangle\simeq \frac{2^{2\nu_0-2}\Gamma^2(\nu_0)}{\pi}a^{-2}(-\eta)\int_{p<Ha} \frac{d^3p}{(2\pi)^3}\ \frac{1}{(-p\eta)^{2\nu_0}}.
\end{align}
As seen in (\ref{nu1}), $\nu_0\ge 3/2$ and therefore the propagator has an IR divergence. 
The upper bound of the momentum integral should be on the horizon since the spectrum is dominant at super-horizon scales. 
To regularize the IR divergence, we introduce an IR cut-off $p_0$ which fixes the minimum value of the comoving momentum. 
In these settings, the IR contribution to the propagator is given by the growing time dependence: 
\begin{align}
\langle\varphi_0^2(x)\rangle
&\simeq\frac{2^{2\nu_0-3}\Gamma^2(\nu_0)}{\pi^3}a^{-2}(-\eta)^{-2}\int^\frac{1}{1-\epsilon_0}_{-p_0\eta}\frac{d(-p\eta)}{(-p\eta)^{2\nu_0-2}} \notag\\
&=\frac{(2-2\epsilon_0)^{2\nu_0-3}\Gamma^2(\nu_0)}{\pi^3}a^{-2}(-\eta)^{-2}\int^1_{-(1-\epsilon_0)p_0\eta}\frac{dx}{x^{2\nu_0-2}} \notag\\
&=\frac{(2-2\epsilon_0)^{2\nu_0-3}\Gamma^2(\nu_0)}{\pi^3}\big\{(1-\epsilon_0)H_0\big\}^2a^{-2\epsilon_0} \notag\\
&\hspace{10em}\times \frac{1-\epsilon_0}{2\epsilon_0}\big\{(a/a_0)^{2\epsilon_0}-1\big\}. 
\label{p1}\end{align}
Here we introduced an initial time $a_0$ which is defined by the IR cut-off: $a_0\equiv (p_0/H_0)^\frac{1}{1-\epsilon_0}$.  

Physically speaking, $a/p_0$ means a size of the Universe and it expands with the scale factor. 
Let us recall that super-horizon scales can be represented by the inequality: 
\begin{align}
a/p_0>1/H\ \Leftrightarrow\ p_0<\dot{a}. 
\label{superhorizon}\end{align}
In accelerating universes $\ddot{a}>0$, the degrees of freedom at super-horizon scales increase with cosmic evolution. 
That is one reason why the propagator acquires the growing time dependence after the momentum integration (\ref{p1}).  

We should also note that if the momentum integration is insensitive to the IR cut-off, 
the time dependence of the propagator is given only by the overall factor: 
\begin{align}
\big\{(1-\epsilon_0)H_0\big\}^2a^{-2\epsilon_0}\propto H^2. 
\end{align}
Namely, if so, the propagator respects the scaling law (\ref{scalinglaw}). 
As seen in (\ref{p1}), the momentum integration is sensitive to the increase of the degrees of freedom at super-horizon scales. 
The scaling law is broken as it is multiplied by the additional time dependence: 
\begin{align}
\frac{1-\epsilon_0}{2\epsilon_0}\big\{(a/a_0)^{2\epsilon_0}-1\big\}. 
\label{additional}\end{align}

Here we mention a massless and non-minimally coupled scalar field. 
Specifically the additional term is introduced to the quadratic action 
\begin{align}
\int\sqrt{-g}\ \big[-\frac{1}{2}\xi R\varphi^2\big]. 
\end{align}
For this scalar field, the index of the Bessel function is deformed as 
\begin{align}
\nu_0=\frac{3-\epsilon_0}{2(1-\epsilon_0)}\ \to\ \sqrt{\frac{(3-\epsilon_0)^2}{4(1-\epsilon_0)^2}-\xi\frac{6(2-\epsilon_0)}{(1-\epsilon_0)^2}}. 
\end{align}
If $\xi$ is large enough, the scaling law is respected since there is no sensitivity of the IR cut-off. 
That is why we consider a massless and minimally coupled scalar field in this paper. 

In interacting field theories with massless and minimally coupled scalar fields, physical quantities may acquire growing time dependences through internal propagators. 
We call them the IR effects in accelerating universes. 
The IR effects at each loop level manifest as polynomials in a factor growing with time. 
At late times, the leading IR effects come from the leading powers of the factor at each loop level.  
This fact indicates that perturbation theories breaks down after a large enough cosmic expansion. 
In the next section, we discuss how to evaluate the IR effects nonperturbatively. 

Before investigating interacting field theories, we show that the propagator (\ref{p1}) can be written as the non-local from: 
\begin{align}
\langle\varphi_0^2(x)\rangle\simeq \frac{(2-2\epsilon_0)^{2\nu_0-3}\Gamma^2(\nu_0)}{\pi^3}(1-\epsilon_0)^3\int^a_{a_0} (d\log a')\ H'^2, 
\label{p1'}\end{align}
where $H'$ denotes $H(t')$. 
Such nonlocal representations are also used in \cite{Starobinsky2009,Starobinsky2010-1,Starobinsky2010-2}.  
At the dS limit $\epsilon_0\to 0$, the integration reproduces the logarithmic dependence of the scale factor which is found in \cite{Vilenkin1982,Linde1982,Starobinsky1982}: 
\begin{align}
\langle\varphi_0^2(x)\rangle\simeq\frac{H_0^2}{4\pi^2}\log (a/a_0). 
\end{align} 
In (\ref{p1'}), the scaling law is respected for the integrand but it is broken after the integration. 
The nonlocal representation is useful when we count the leading powers of the growing time dependence. 

\section{Nonperturbative analyses of IR effects in an accelerating universe with $\dot{\epsilon}=0$}\label{NC}
\setcounter{equation}{0}

\subsection{Resummation formula of leading IR effects}\label{NC1}

In investigating the IR effects nonperturbatively, a rational approach is to resum the leading IR effects to all loop orders. 
The resummation formula in dS space has been derived by N. C. Tsamis and R. P. Woodard \cite{Woodard2005}. 
In this subsection, we extend the nonperturbative investigation in an accelerating universe whose slow-roll parameter is constant. 

In this paper, we consider an interacting field theory whose nonlinear term is given by a potential 
\begin{align}
S=\int\sqrt{-g}d^4x\ \big[-\frac{1}{2}g^{\mu\nu}\partial_\mu \varphi\partial_\nu\varphi-V(\varphi)\big]. 
\end{align}
As a starting point, we consider the Yang-Feldman formalism: 
\begin{align}
\varphi(x)=\varphi_0(x)-i\int\sqrt{-g'}d^4x'\ G^R(x,x')\frac{\partial}{\partial \varphi}V(\varphi(x')), 
\label{YF}\end{align}
where $G^R$ is the retarded propagator: 
\begin{align}
G^R(x,x')=\theta(t-t')[\varphi_0(x),\varphi_0(x')]. 
\end{align}
The Yang-Feldman formalism satisfies the equation of motion and describes quantum effects from all regions of momentum scale. 
Our purpose is to derive the equation which is valid up to the leading IR effects. 

For the linear term in the right-hand side of (\ref{YF}), we extract the leading IR behavior as 
\begin{align}
\varphi_0(x)\simeq\bar{\varphi}_0(x)
&\equiv\int\frac{d^3p}{(2\pi)^3}\ \theta(Ha-p) 
\big[-i\frac{2^{\nu_0-1}\Gamma(\nu_0)}{\sqrt{\pi}}\frac{a^{-1}(-\eta)^\frac{1}{2}}{(-p\eta)^{\nu_0}}e^{+i{\bf p}\cdot{\bf x}}a_{\bf p}+\text{(h.c.)}\big] \notag\\
&=\int\frac{d^3p}{(2\pi)^3}\ \theta(Ha-p) 
\big[-i\frac{2^{\nu_0-1}\Gamma(\nu_0)}{\sqrt{\pi}}\frac{\big\{(1-\epsilon_0)H_0\big\}^\frac{1}{1-\epsilon_0}}{p^{\nu_0}}e^{+i{\bf p}\cdot{\bf x}}a_{\bf p}+\text{(h.c.)}\big]. 
\label{linear1}\end{align}
Here we introduced the step function which is nonzero at super-horizon scales. 
The leading IR term of the wave function has no time dependence since quantum fluctuations of a massless and minimally coupled scalar field are frozen at super-horizon scales. 
As seen in (\ref{w1}) and (\ref{p1}), the leading IR behavior of the linear term induces the growing time dependences to the Wightman functions: 
\begin{align}
\langle\varphi_0(x)\varphi_0(x')\rangle,\hspace{1em}\langle\varphi_0(x')\varphi_0(x)\rangle. 
\end{align}

In the Schwinger-Keldysh formalism \cite{Schwinger1961,Keldysh1964}, perturbation theories consist not only of the Wightman functions but also of the retarded propagator. 
To be more precise, each vertex integral contains one retarded propagator due to the causality. 
The nonlinear term in the right-hand side of (\ref{YF}) corresponds to the vertex integral. 

Let us extract the leading IR behavior of the retarded propagator in the nonlinear term. 
In evaluating the retarded propagator, we need to know the real and the imaginary parts of the wave function. 
From (\ref{ex1}) and (\ref{ex2}), the leading IR terms of the imaginary and the real parts come from the zeroth orders of the expansions of $J_{-\nu_0}$ and $J_{\nu_0}$ respectively: 
\begin{align}
\phi_{\bf p}(x)&\simeq\big[-i\frac{2^{\nu_0-1}\Gamma(\nu_0)}{\sqrt{\pi}}\frac{a^{-1}(-\eta)^\frac{1}{2}}{(-p\eta)^{\nu_0}} 
+\frac{\sqrt{\pi}}{2^{\nu_0+1}\nu_0\Gamma(\nu_0)}a^{-1}(-\eta)^\frac{1}{2}(-p\eta)^{\nu_0}\big]e^{+i{\bf p}\cdot{\bf x}} \notag\\
&=\big[-i\frac{2^{\nu_0-1}\Gamma(\nu_0)}{\sqrt{\pi}}\frac{\big\{(1-\epsilon_0)H_0\big\}^\frac{1}{1-\epsilon_0}}{p^{\nu_0}}
+\frac{\sqrt{\pi}}{2^{\nu_0+1}\nu_0\Gamma(\nu_0)}\frac{p^{\nu_0}}{\big\{(1-\epsilon_0)H_0\big\}^{\frac{1}{1-\epsilon_0}+1}}a^{-(3-\epsilon_0)}\big]e^{+i{\bf p}\cdot{\bf x}}. 
\end{align}
The leading IR behavior of the retarded propagator is given by the cross terms between the above two parts. 
The contribution is spatially local in the sense that its spectrum has no momentum dependence and then it is proportional to the spatial delta function 
\begin{align}
G^R(x,x')&\simeq  -\frac{i}{2\nu_0(1-\epsilon_0)H_0}\ \theta(t-t')\int \frac{d^3p}{(2\pi)^3}\ \big[a'^{-(3-\epsilon_0)}-a^{-(3-\epsilon_0)}\big]e^{+i{\bf p}\cdot({\bf x}-{\bf x}')} \notag\\
&=-\frac{i}{(3-\epsilon_0)H_0}\ \theta(t-t') \big[a'^{-(3-\epsilon_0)}-a^{-(3-\epsilon_0)}\big]\delta^{(3)}({\bf x}-{\bf x}'). 
\label{retarded1}\end{align}

Substituting the leading IR behavior of the retarded propagator, the nonlinear term is written as 
\begin{align}
&-i\int\sqrt{-g'}d^4x'\ G^R(x,x')\frac{\partial}{\partial \varphi}V(\varphi(x')) \notag\\
\simeq&-\frac{1}{3-\epsilon_0}\int^tdt'\ H_0^{-1}a'^{\epsilon_0}\big[1-(a'/a)^{3-\epsilon_0}\big]\frac{\partial}{\partial \varphi}V(\varphi(t',{\bf x})). 
\end{align}
The integration of the second term does not contribute to the leading IR effects since $(a'/a)^{3-\epsilon_0}\ll 1$ at the far past $t'\ll t$. 
Neglecting the second term, the vertex integral is evaluated as 
\begin{align}
-i\int\sqrt{-g'}d^4x'\ G^R(x,x')\frac{\partial}{\partial \varphi}V(\varphi(x')) 
\simeq-\frac{1}{3-\epsilon_0}\int^tdt'\ H'^{-1}\frac{\partial}{\partial \varphi}V(\varphi(t',{\bf x})). 
\label{nonlinear1}\end{align}
Note that the time dependence of the integrand can be expressed by $H'$. 
In a similar way to (\ref{p1'}), it is useful to represent the integral in terms of $\log a'$ as 
\begin{align}
-\frac{1}{3-\epsilon_0}\int^ad(\log a')\ H'^{-2}\frac{\partial}{\partial \varphi}V(\varphi(t',{\bf x})). 
\label{nonlinear1'}\end{align}
As seen in (\ref{retarded1}) and (\ref{nonlinear1'}), the retarded propagator has no IR divergence in itself 
while the vertex integration of it induces the growing time dependence. 

For example in $\varphi^4$ theory, two additional propagators appear with the loop level is increased by one. 
One of them is the retarded propagator and the other is the Wightman function. 
From (\ref{p1'}) and (\ref{nonlinear1'}), the leading IR effects are given by infinite series of the following factor: 
\begin{align}
\lambda \int d(\log a')\ H'^{-2}\int d(\log a'')\ H''^2,  
\label{counting}\end{align}
where $\lambda$ denotes a coupling constant. 
The amplitude of this factor grows with cosmic evolution. 
Therefore, even if the couping constant is small $\lambda\ll 1$, the perturbation theory breaks down after an enough time passed: $\lambda \int d(\log a')\ H'^{-2}\int d(\log a'')\ H''^2\sim 1$. 

Let us derive the equation which describes the leading IR effects nonperturbatively. 
Applying the approximations (\ref{linear1}) and (\ref{nonlinear1'}), the Yang-Feldman formalism is reduced as 
\begin{align}
\varphi(x)=\bar{\varphi}_0(x)-\frac{1}{3-\epsilon_0}\int^t dt'\ H'^{-1}\frac{\partial}{\partial\varphi}V(\varphi(t',{\bf x})).
\label{RYF}\end{align}
We should note that the time dependence of the nonlinear term appears only in the upper bound of the integral. 
Differentiating both sides of (\ref{RYF}) with respect to $t$, we obtain the local equation: 
\begin{align}
\dot{\varphi}(x)=\dot{\bar{\varphi}}_0(x)-\frac{1}{(3-\epsilon_0)H}\frac{\partial}{\partial\varphi}V(\varphi(t,{\bf x})). 
\label{langevin1}\end{align}
As discussed below (\ref{linear1}), the leading IR term of the wave function is constant and so the time dependence of $\bar{\varphi}_0$ appears only through the step function which denotes super-horizon scales. 
As a direct consequence of this fact, the correlation function of $\dot{\bar{\varphi}}_0$ is proportional to the delta function 
\begin{align}
\langle \dot{\bar{\varphi}}_0(t,{\bf x})\dot{\bar{\varphi}}_0(t',{\bf x})\rangle
&=\int\frac{d^3p}{(2\pi)^3}\ \big\{(1-\epsilon_0)H^2a\big\}^2\delta(Ha-p)\delta(Ha-H'a')\notag\\
&\hspace{5.2em}\times \frac{2^{2\nu_0-2}\Gamma^2(\nu_0)}{\pi}\frac{\big\{(1-\epsilon_0)H_0\big\}^\frac{2}{1-\epsilon_0}}{p^{2\nu_0}}\notag\\
&=\frac{(2-2\epsilon_0)^{2\nu_0-3}\Gamma^2(\nu_0)}{\pi^3}\big\{(1-\epsilon_0)H\big\}^3\delta (t-t').
\label{noise1}\end{align}
We call this type of fluctuation a white noise. 
The equation (\ref{langevin1})-(\ref{noise1}) is known as a Langevin equation with a white noise \cite{Risken}\footnote{The standard Langevin equation describes the time evolution of the position of a particle. Here a coordinate space is replaced by a field space. }.

The Langevin equation can be translated to the equation of the probability density $\rho$, which is called the Fokker-Planck equation: 
\begin{align}
\dot{\rho}(t,\phi)=\frac{1}{2}A_0H^3\frac{\partial^2}{\partial\phi^2}\rho(t,\phi)+\frac{1}{(3-\epsilon_0)H}\frac{\partial}{\partial\phi}\Big(\rho(t,\phi)\frac{\partial}{\partial\phi}V(\phi)\Big), 
\label{FP1}\end{align}
\begin{align}
A_0=\frac{(2-2\epsilon_0)^{2\nu_0-3}\Gamma^2(\nu_0)}{\pi^3}(1-\epsilon_0)^3.
\label{A1}\end{align}
Using the probability density, the vev of any field operator is given by  
\begin{align}
\langle F(\varphi(x))\rangle=\int^\infty_{-\infty}d\phi\ \rho(t,\phi)F(\phi),\hspace{1em}F\text{: any function}. 
\label{probability}\end{align}
Note that $\phi$ is not an operator but a $c$-number. 

In the free field theory $V=0$, the solution of the Fokker-Planck equation is given by the Gaussian distribution with the growing variance: 
\begin{align}
\rho(t,\phi)=\frac{1}{\sqrt{2\pi\sigma^2}}\exp\Big(-\frac{1}{2}\frac{\phi^2}{\sigma^2}\Big), 
\end{align}
\begin{align}
\sigma^2=A_0\int^t_{t_0}dt'\ H'^3 =\frac{(2-2\epsilon_0)^{2\nu_0-3}\Gamma^2(\nu_0)}{\pi^3}(1-\epsilon_0)^3\int^a_{a_0} (d\log a')\ H'^2. 
\end{align}
The probability density reproduces the IR behavior of the propagator (\ref{p1'}). 
The IR cut-off dependence appears as an initial time. 

In interacting field theories, the most interesting objects are eventual contributions to physical quantities. 
In order to evaluate them, we need to solve the Fokker-Planck equation for a final state. 
If we know the final state solution $\rho_\infty$, the eventual contribution to the vev of any field operator can be evaluated by substituting it in the integral (\ref{probability}).  

In dS space $\epsilon_0= 0$, we can obtain an analytical solution of the final state for an arbitrary stable potential. 
Assuming that an equilibrium state is eventually established: 
\begin{align}
\rho(t,\phi)\to\rho_\infty(\phi), 
\end{align}
the Fokker-Planck equation (\ref{FP1})-(\ref{A1}) is written as 
\begin{align}
0=\frac{H_0^3}{8\pi^2}\frac{\partial^2}{\partial\phi^2}\rho_\infty(\phi)+\frac{1}{3H_0}\frac{\partial}{\partial\phi}\Big(\rho_\infty(\phi)\frac{\partial}{\partial\phi}V(\phi)\Big). 
\end{align} 
The final state solution is given by 
\begin{align}
\rho_\infty(\phi)&=N^{-1}\exp\Big(-\frac{8\pi^2}{3H_0^4}V(\phi)\Big), \notag\\
N&=\int^\infty_{-\infty}d\phi\ \exp\Big(-\frac{8\pi^2}{3H_0^4}V(\phi)\Big). 
\end{align}
For example in $\varphi^4$ theory $V=\frac{\lambda}{4!}\varphi^4$, the vev of the potential approaches the saturation value:  
\begin{align}
\langle V(\varphi(x))\rangle_\infty=\frac{3H_0^4}{32\pi^2}. 
\end{align} 
The saturation value is not suppressed by $\lambda\ll 1$ and so it is surely a nonperturbative IR effect. 

On the background with $\epsilon_0\not=0$, the time independence of the final state solution does not hold true since the Hubble parameter is time dependent. 
For a general potential, we cannot obtain an analytical solution of the final state. 
T. Prokopec found that we can obtain it if the potential is written as the following form \cite{Prokopec2015-2}:  
\begin{align}
V(\varphi(x))=H^4\tilde{V}(\tilde{\varphi}(x)),\hspace{1em}\tilde{\varphi}(x)\equiv \frac{\varphi(x)}{H}. 
\label{dimensionless}\end{align}
In such a model, adapting the ansatz that the scaling law is eventually restored: 
\begin{align}
\rho(t,\phi)\to \rho_\infty(t,\phi)=\frac{1}{H}\tilde{\rho}_\infty(\tilde{\phi}),\hspace{1em}\tilde{\phi}\equiv \frac{\phi}{H}, 
\end{align}
the Fokker-Planck equation (\ref{FP1})-(\ref{A1}) is written as 
\begin{align}
\epsilon_0\frac{\partial}{\partial\tilde{\phi}}\Big(\tilde{\rho}_\infty(\tilde{\phi})\ \tilde{\phi}\Big)
=\frac{1}{2}A_0\frac{\partial^2}{\partial\tilde{\phi}^2}\tilde{\rho}_\infty(\tilde{\phi})+\frac{1}{3-\epsilon_0}\frac{\partial}{\partial\tilde{\phi}}\Big(\tilde{\rho}_\infty(\tilde{\phi})\frac{\partial}{\partial\tilde{\phi}}\tilde{V}(\tilde{\phi})\Big). 
\end{align} 
The final state solution is given by  
\begin{align}
\tilde{\rho}_\infty(\tilde{\phi})&=\tilde{N}^{-1}\exp\Big(+\frac{\epsilon_0}{A_0}\tilde{\phi}^2-\frac{2}{(3-\epsilon_0)A_0}\tilde{V}(\tilde{\phi})\Big), \notag\\
\tilde{N}&=\int^\infty_{-\infty}d\tilde{\phi}\ \exp\Big(+\frac{\epsilon_0}{A_0}\tilde{\phi}^2-\frac{2}{(3-\epsilon_0)A_0}\tilde{V}(\tilde{\phi})\Big). 
\label{solution}\end{align}
Here the tachyonic term appears due to the existence of a nonzero slow-roll parameter. 

As an example, we consider $\varphi^4$ theory which belongs to the solvable case (\ref{dimensionless}). 
Since the exponent of (\ref{solution}) is not written as a monomial, the evaluation of the integral (\ref{probability}) is complicated. 
Specifically, the eventual contribution to the vev of the potential is given by 
\begin{align}
\langle V(\varphi(x))\rangle_\infty&=\frac{(3-\epsilon_0)A_0}{2}H^4
\frac{\sum^\infty_{n=0}\frac{1}{n!}\Gamma(\frac{n}{2}+\frac{5}{4})\big(\sqrt{\frac{12(3-\epsilon_0)}{A_0}}\frac{\epsilon_0}{\lambda^\frac{1}{2}}\big)^n}
{\sum^\infty_{n=0}\frac{1}{n!}\Gamma(\frac{n}{2}+\frac{1}{4})\big(\sqrt{\frac{12(3-\epsilon_0)}{A_0}}\frac{\epsilon_0}{\lambda^\frac{1}{2}}\big)^n} \notag\\
&=H^4\big[\frac{(3-\epsilon_0)A_0}{8}+\mathcal{O}(\epsilon_0/\lambda^\frac{1}{2})\big]. 
\label{V}\end{align}
The emergence of $\epsilon_0/\lambda^\frac{1}{2}$ can be explained as follows. 
Let us recall that the leading IR effects are expressed as infinite series of the factor (\ref{counting}). 
This factor can be rewritten as 
\begin{align}
\int d(\lambda^\frac{1}{2}\log a')\ H'^{-2}\int d(\lambda^\frac{1}{2}\log a'')\ H''^2, 
\end{align}
and the Hubble parameter is a function of $\epsilon_0\log a$: $H=H_0\exp(-\epsilon_0\log a)$. 
Therefore, if $\epsilon_0\ll \lambda^\frac{1}{2}$, the Hubble parameter can be moved outside of the integral just like in dS space: 
\begin{align}
H^{-2}H^2\int d(\lambda^\frac{1}{2}\log a')\int d(\lambda^\frac{1}{2}\log a'')\sim \lambda\log^2(a/a_0). 
\label{approximation}\end{align}
The higher orders in (\ref{V}) mean the corrections to this evaluation. 

In an accelerating universe whose slow-roll parameter is constant, 
we derived the Langevin equation from the Yang-Feldman formalism by extracting the leading IR effects. 
In the derivation process, the loop level and the type of Feynman diagram are not specified. 
Therefore, the stochastic equation describes the leading IR effects nonperturbatively. 
Especially, the final state solution of the Fokker-Planck equation is useful in evaluating eventual contributions to physical quantities.  

\subsection{Semiclassical description of scalar fields}\label{NC2}

Before N. C. Tsamis and R. P. Woodard derived the Langevin equation in dS space by the resummation of the leading IR effects, 
A. A. Starobinsky and J. Yokoyama derived the same equation from a semiclassical view point \cite{Starobinsky1986,Starobinsky1994}. 
Here we confirm that after a slight modification, the semiclassical approach is consistent with the resummation formula also in an accelerating universe whose slow-roll parameter is constant.    

First of all, we make a naive extension of the previous studies \cite{Starobinsky1986,Starobinsky1994}. 
The naive investigation leads to an inconsistency with the resummation formula.
We discuss how to resolve the inconsistency in the latter part of this subsection. 

We start with the equation of motion: 
\begin{align}
\Big(\frac{\partial^2}{\partial t^2}+3H\frac{\partial}{\partial t}-\frac{1}{a^2}\frac{\partial^2}{\partial{\bf x}^2}\Big)\varphi(x)=-\frac{\partial}{\partial\varphi}V(\varphi(x)). 
\label{EOM}\end{align}
From the second order differential equation, we extract the equation which describes the IR dynamics at super-horizon scales. 
Since our interest is about physics at late times, we neglect the second derivative terms of (\ref{EOM}) naively as 
\begin{align}
3H\frac{\partial}{\partial t}\varphi(x)=-\frac{\partial}{\partial\varphi}V(\varphi(x)).
\label{neglect0}\end{align}

Let us recall that in accelerating universes, more degrees of freedom go outside of the horizon $P=H$ with cosmic evolution.  
In other words, super-horizon modes are sourced by sub-horizon modes. 
Based on this perspective, we divide the scalar field into super-horizon modes $\bar{\varphi}$ and sub-horizon modes $\varphi_\text{UV}$:  
\begin{align}
\varphi(x)=\bar{\varphi}(x)+\varphi_\text{UV}(x), 
\end{align}
and identify $\varphi_\text{UV}$ as a source of $\bar{\varphi}$.   
To be more precise, we set $\varphi_\text{UV}$ as 
\begin{align}
\varphi_\text{UV}(x)
&\equiv\int\frac{d^3p}{(2\pi)^3}\ \theta(p-Ha) 
\big[-i\frac{2^{\nu_0-1}\Gamma(\nu_0)}{\sqrt{\pi}}\frac{\big\{(1-\epsilon_0)H_0\big\}^\frac{1}{1-\epsilon_0}}{p^{\nu_0}}e^{+i{\bf p}\cdot{\bf x}}a_{\bf p}+\text{(h.c.)}\big], 
\label{UV1}\end{align}
and keep the time derivative of it in the equation (\ref{neglect0}): 
\begin{align}
3H\frac{\partial}{\partial t}\big\{\bar{\varphi}(x)+\varphi_\text{UV}(x)\big\}=-\frac{\partial}{\partial\bar{\varphi}}V(\bar{\varphi}(x)), 
\end{align}

In the setting (\ref{UV1}), the time derivative of $\varphi_\text{UV}$ is equal to that of (\ref{linear1}) except for the sign: 
\begin{align}
\dot{\varphi}_\text{UV}(x)=-\dot{\bar{\varphi}}_0(x). 
\end{align}
By moving it from the left-hand side to the right-hand side, we obtain the following Langevin equation: 
\begin{align}
\dot{\bar{\varphi}}(x)=\dot{\bar{\varphi}}_0(x)-\frac{1}{3H}\frac{\partial}{\partial\bar{\varphi}}V(\bar{\varphi}(t,{\bf x})). 
\end{align}
The noise term is consistent with the result obtained by the resummation formula (\ref{langevin1})-(\ref{noise1}). 
In contrast, the coefficient of the nonlinear term is inconsistent with it except in dS space $\epsilon_0= 0$. 
As discussed below, the difference of the coefficient originates in the choice of the time coordinate the second derivative with respect to which we neglect. 

The consistency with the resummation formula is obtained in the following way. 
In terms of the e-folding number: $N\equiv \log a$, the equation of motion is written as 
\begin{align}
\Big(H^2\frac{\partial^2}{\partial N^2}+(3-\epsilon_0)H^2\frac{\partial}{\partial N}-\frac{1}{a^2}\frac{\partial^2}{\partial{\bf x}^2}\Big)\varphi(x)=-\frac{\partial}{\partial\varphi}V(\varphi(x)). 
\end{align}
Neglecting the second derivatives with respect to $N$, ${\bf x}$ rather than with respect to $t$, ${\bf x}$, 
we obtain the equation different from (\ref{neglect0}): 
\begin{align}
(3-\epsilon_0)H^2\frac{\partial}{\partial N}\varphi(x)=(3-\epsilon_0)H\frac{\partial}{\partial t}\varphi(x)=-\frac{\partial}{\partial\varphi}V(\varphi(x)). 
\label{neglect1}\end{align}

Taking account of the contribution from sub-horizon scales in the same way as (\ref{UV1}),  
\begin{align}
(3-\epsilon_0)H\frac{\partial}{\partial t}\big\{\bar{\varphi}(x)+\varphi_\text{UV}(x)\big\}=-\frac{\partial}{\partial\bar{\varphi}}V(\bar{\varphi}(x)), 
\end{align}
we obtain the Langevin equation which agrees with the result in the previous subsection (\ref{langevin1})-(\ref{noise1}): 
\begin{align}
\dot{\bar{\varphi}}(x)=\dot{\bar{\varphi}}_0(x)-\frac{1}{(3-\epsilon_0)H}\frac{\partial}{\partial\bar{\varphi}}V(\bar{\varphi}(t,{\bf x})). 
\label{same1}\end{align}  

In an accelerating universe whose slow-roll parameter is constant, 
we find that the semiclassical description of the scalar field leads to the same Langevin equation, which is derived by the resummation formula. 
In the derivation process, it is crucial to neglect the second derivative with respect to the e-folding number. 

Actually if the slow-roll parameter is time dependent, the e-folding number is no longer an appropriate choice to obtain the consistency with the resummation formula.   
In Subsection \ref{NG2}, we discuss which choice of the time coordinate is consistent with the resummation of the leading IR effects in a general accelerating universe. 

\section{Free scalar field theory in a general accelerating universe}\label{FG}
\setcounter{equation}{0}

The degrees of freedom at super-horizon scales increase with time as far as the cosmic expansion is accelerated. 
This fact indicates that physical quantities may acquire growing time dependences through quantum fluctuations at super-horizon scales even if the slow-roll parameter is time dependent $\dot{\epsilon}\not=0$. 
In this section, we investigate the IR effects in a general accelerating universe. 

On a general background whose metric is given by (\ref{metric}), the wave function for a massless and minimally coupled scalar field has been derived \cite{Woodard2003}. 
By reference to the previous study, the leading IR term of the wave function is evaluated as\footnote{For simplicity, the reference time $t_i$ in \cite{Woodard2003} is set at the time of  horizon crossing $t_i$ in this paper. } 
\begin{align}
\phi_{\bf p}(x)\simeq -i\frac{2^{\nu_*-1}\Gamma(\nu_*)}{\sqrt{\pi}}\frac{\big\{(1-\epsilon_*)H_*a^{\epsilon_*}\big\}^\frac{1}{1-\epsilon_*}}{p^{\nu_*}} e^{+i{\bf p}\cdot{\bf x}}, 
\label{w2}\end{align}
where the subscript $*$ denotes the time of horizon crossing: $H_*a_*\equiv p$ and $\nu$ is defined in the same way as (\ref{nu1}) 
\begin{align}
\nu=\frac{3-\epsilon}{2(1-\epsilon)}=\frac{3}{2}+\frac{\epsilon}{1-\epsilon}. 
\end{align}
Since each parameter in (\ref{w2}) is expressed at the time of horizon crossing, the leading IR term of the wave function depends on $p$ rather than on $t$. 
This fact is consistent with the freezing of quantum fluctuations at super-horizon scales. 

From (\ref{w2}), the IR behavior of the propagator is given by 
\begin{align}
\langle\bar{\varphi}_0^2(x)\rangle
&\simeq\int^{Ha}_{p_0}dp\ \frac{2^{2\nu_*-3}\Gamma^2(\nu_*)}{\pi^3}\frac{\big\{(1-\epsilon_*)H_*a^{\epsilon_*}\big\}^\frac{1}{1-\epsilon_*}}{p^{2\nu_*-2}} \notag\\
&=\int^a_{a_0} d(\log a')\ (1-\epsilon')\times \frac{(2-2\epsilon')^{2\nu'-3}\Gamma^2(\nu')}{\pi^3}\big\{(1-\epsilon')H'\big\}^2. 
\label{p2}\end{align}
In the second line, the integral variable is changed as $p=H'a'$. 
We should note that the integral measure induces $(1-\epsilon')$ as follows 
\begin{align}
dp=(1-\epsilon')pd(\log a'). 
\end{align}
Therefore, the propagator is growing with time as far as accelerated expansion continues. 
Furthermore it should be emphasized that the nonlocal quantum effects induced by the propagator cannot be subtracted by local counter terms. 

\section{Nonperturbative analyses of IR effects in a general accelerating universe}\label{NG}
\setcounter{equation}{0}

\subsection{Resummation formula of leading IR effects}\label{NG1}

The growing time dependence of the free propagator (\ref{p2}) indicates that the breakdown of perturbation theories takes place in a general accelerating universe.    
We extend the resummation formula in Subsection \ref{NC1} as it is valid even if the slow-roll parameter is time dependent.  

To begin with, we extract the leading IR behavior of the free scalar field as 
\begin{align}
\varphi_0(x)\simeq\bar{\varphi}_0(x)\equiv &\int \frac{d^3p}{(2\pi)^3}\ \theta(Ha-p)
\big[-i\frac{2^{\nu_*-1}\Gamma(\nu_*)}{\sqrt{\pi}}\frac{\big\{(1-\epsilon_*)H_*a^{\epsilon_*}\big\}^\frac{1}{1-\epsilon_*}}{p^{\nu_*}}e^{+i{\bf p}\cdot{\bf x}}a_{\bf p}+\text{(h.c.)}\big]. 
\label{linear2}\end{align}
In a similar way to (\ref{linear1}), the time dependence of $\bar{\varphi}_0$ appears only through the step function which denotes super-horizon scales.  

As a next step, we extract the leading IR behavior of the retarded propagator. 
By reference to \cite{Woodard2003}, the leading IR terms of the imaginary and real parts of the wave function are given by 
\begin{align}
\phi_{\bf p}(x)\simeq \Big[&-i\frac{2^{\nu_*-1}\Gamma(\nu_*)}{\sqrt{\pi}}\frac{\big\{(1-\epsilon_*)H_*a^{\epsilon_*}\big\}^\frac{1}{1-\epsilon_*}}{p^{\nu_*}} \notag\\
&+\frac{\sqrt{\pi}}{2^{\nu_*}\Gamma(\nu_*)}\frac{p^{\nu_*}}{\big\{(1-\epsilon_*)H_*a^{\epsilon_*}\big\}^\frac{1}{1-\epsilon_*}}
\big\{\int^{t_*}_tdt'\ a'^{-3}+\frac{1}{3-\epsilon_*}a_*^{-3}H_*^{-1}\big\}\Big]e^{+i{\bf p}\cdot{\bf x}}. 
\end{align}
The time dependence of the IR behavior is expressed by the linear combination of $1$ and $\int^t dt'\ a'^{-3}$: 
independent solutions of the equation of motion at the limit $p\to 0$. 
From the above, the retarded propagator is evaluated as 
\begin{align}
G^R(x,x')&\simeq -i\theta(t-t')\int\frac{d^3p}{(2\pi)^3}\ \big[\int^t_{t'}dt''\ {a''}^{-3}\ \big]e^{+i{\bf p}\cdot({\bf x}-{\bf x}')} \notag\\
&=-i\theta(t-t')\big[\int^t_{t'}dt''\ {a''}^{-3}\ \big] \delta^{(3)}({\bf x}-{\bf x}'). 
\label{retarded2}\end{align}
In a similar way to (\ref{retarded1}), the retarded propagator is proportional to the spatial delta function up to the leading IR behavior. 

Substituting (\ref{retarded2}), the vertex integral is written as 
\begin{align}
&-i\int \sqrt{-g'}d^4x'\ G^R(x,x')\frac{\partial}{\partial\varphi}V(\varphi(x')) \notag\\
\simeq&-\int^t dt'\ {a'}^3\big[\int^t_{t'}dt''\ {a''}^{-3}\ \big]\frac{\partial}{\partial\varphi}V(\varphi(t',{\bf x})) \notag\\
=&-\int^t dt'\ {a'}^3\big[\int^\infty_{t'}dt''\ {a''}^{-3}-\int^\infty_tdt''\ {a''}^{-3}\ \big]\frac{\partial}{\partial\varphi}V(\varphi(t',{\bf x})) \notag\\
=&-\int^t dt'\ \big({a'}^3\int^\infty_{t'}dt''\ {a''}^{-3}\big) \Big[1-\frac{\int^\infty_tdt''\ {a''}^{-3}}{\int^\infty_{t'}dt''\ {a''}^{-3}}\Big]\frac{\partial}{\partial\varphi}V(\varphi(t',{\bf x})).
\end{align}
We should note that the amplitude of $\int^\infty_tdt''\ {a''}^{-3}$ monotonically decreases with time 
and so $\int^\infty_tdt''\ {a''}^{-3}/\int^\infty_{t'}dt''\ {a''}^{-3}\ll 1$ at the far past $t'\ll t$.   
Therefore, the integration of the second term is negligible up to the leading IR effects: 
\begin{align}
-i\int \sqrt{-g'}d^4x'\ G^R(x,x')\frac{\partial}{\partial\varphi}V(\varphi(x')) 
\simeq-\int^t dt'\ \big({a'}^3\int^\infty_{t'}dt''\ {a''}^{-3}\big) \frac{\partial}{\partial\varphi}V(\varphi(t',{\bf x})). 
\label{nonlinear2}\end{align}
Note that the time dependence of the integral appears only in its upper bound. 

Applying the approximations (\ref{linear2}) and (\ref{nonlinear2}), and then differentiating both sides with respect to $t$, 
the Yang-Feldman formalism (\ref{YF}) is reduced to the Langevin equation: 
\begin{align}
\dot{\varphi}(x)=\dot{\bar{\varphi}}_0(x)-\big(a^3\int^\infty_{t}dt'\ {a'}^{-3}\big)\frac{\partial}{\partial\varphi}V(\varphi(x)), 
\label{langevin2}\end{align}
where $\dot{\bar{\varphi}}_0$ is given by the white noise due to the freezing of the IR spectrum:  
\begin{align}
\langle\dot{\bar{\varphi}}_0(t,{\bf x})\dot{\bar{\varphi}}_0(t',{\bf x})\rangle&=\int \frac{d^3p}{(2\pi)^3}\ \big\{(1-\epsilon)H^2a\big\}^2\delta(Ha-p)\delta(Ha-H'a') \notag\\
&\hspace{5.2em}\times\frac{2^{2\nu_*-2}\Gamma^2(\nu_*)}{\pi}\frac{\big\{(1-\epsilon_*)H_*a^{\epsilon_*}\big\}^\frac{2}{1-\epsilon_*}}{p^{2\nu_*}} \notag\\
&=\frac{(2-2\epsilon)^{2\nu-3}\Gamma^2(\nu)}{\pi^3}\big\{(1-\epsilon)H\big\}^{3}\delta(t-t'). 
\label{noise2}\end{align}

The Langevin equation (\ref{langevin2})-(\ref{noise2}) can be translated to the following Fokker-Planck equation: 
\begin{align}
\dot{\rho}(t,\phi)=\frac{1}{2}AH^3\frac{\partial^2}{\partial\phi^2}\rho(t,\phi)+\big(a^3\int^\infty_{t}dt'\ {a'}^{-3}\big)\frac{\partial}{\partial\phi}\Big(\rho(t,\phi)\frac{\partial}{\partial\phi}V(\phi)\Big). 
\label{FP2}\end{align}
\begin{align}
A=\frac{(2-2\epsilon)^{2\nu-3}\Gamma^2(\nu)}{\pi^3}(1-\epsilon)^3. 
\label{A2}\end{align}
The coefficient of the second derivative term is positive during the cosmic expansion is accelerated. That is why the propagator grows with time. 
In the free field theory, the variance of the Gaussian distribution: $\sigma^2=\int^t_{t_0}dt'\ A'H'^3$ reproduces the time dependence of the propagator (\ref{p2}).  

On a general background with $\dot{\epsilon}\not=0$, we cannot obtain an exact solution of the final state even in the model with dimensionless couplings (\ref{dimensionless}). 
In a similar way to (\ref{approximation}), if the time variations of $H$, $\epsilon$ are negligible during the IR effects grow, 
we can evaluate the final state solution just like in dS space: 
\begin{align}
\rho_\infty(t,\phi)&\sim N^{-1}\exp\Big(-\frac{2}{A}\big(a^3H\int^\infty_ad(\log a')\ a'^{-3}H'\big)\frac{V(\phi)}{H^4}\Big) \notag\\
N&\sim\int^\infty_{-\infty}d\phi\ \exp\Big(-\frac{2}{A}\big(a^3H\int^\infty_ad(\log a')\ a'^{-3}H'\big)\frac{V(\phi)}{H^4}\Big). 
\label{approximate}\end{align}
In such a situation, the eventual contribution to the vev of the potential in $\varphi^4$ theory is evaluated as 
\begin{align}
\langle V(\varphi(x))\rangle_\infty\sim \frac{A}{8}\big(a^3H\int^\infty_ad(\log a')\ a'^{-3}H'^{-1}\big)^{-1}H^4. 
\end{align}

\subsection{Semiclassical description of scalar fields}\label{NG2}

Here we make the semiclassical derivation of the Langevin equation in a general accelerating universe. 
If we neglect the second derivative with respect to the e-folding number in the same way as (\ref{neglect1}), 
the coefficient of its nonlinear term is given by the replacement $\epsilon_0\to\epsilon$ in (\ref{same1}). 
However the result of the resummation derivation (\ref{langevin2}) shows the different coefficient:    
\begin{align}
a^3\int^\infty_t dt'\ a'^{-3}=\frac{1}{(3-\epsilon)H}+a^3\int^\infty_tdt'\ \frac{\dot{\epsilon}'a'^{-3}}{(3-\epsilon')^2H'}. 
\end{align}
From this fact, we find that the e-folding number is not an appropriate choice of time coordinate for $\dot{\epsilon}\not=0$.  

To obtain the consistency with the resummation formula, we extend the semiclassical description of the scalar field in the following way.  
For a general choice of time coordinate: 
\begin{align}
dT=\mathcal{H}dt, 
\end{align}
the temporal part of d'Alembert operator is written as 
\begin{align}
\frac{\partial}{\partial t^2}+3H\frac{\partial}{\partial t}=\mathcal{H}^2\Big(\frac{\partial^2}{\partial T^2}+\frac{\dot{\mathcal{H}}+3H\mathcal{H}}{\mathcal{H}^2}\frac{\partial}{\partial T}\Big). 
\end{align}
In order to compare $\frac{\partial}{\partial T}\varphi$ directly with $\frac{\partial^2}{\partial T^2}\varphi$, we choose the time coordinate $T$ as its friction coefficient is constant 
\begin{align}
\frac{\dot{\mathcal{H}}+3H\mathcal{H}}{\mathcal{H}^2}=\mu_0\text{ : const.}
\label{friction}\end{align}
The solution of this equation is given by 
\begin{align}
\mathcal{H}=\frac{1}{\mu_0}\big(a^3\int^\infty_tdt'\ a'^{-3}\big)^{-1}. 
\end{align}
Here the boundary condition is set as the amplitude of $T$ monotonically increases with time. 
At the limit $\dot{\epsilon}\to 0$, the time coordinate corresponds to the e-folding number up to the constant scale factor: $dT=\frac{3-\epsilon_0}{\mu_0}dN$. 

Neglecting the second derivatives with respect to $T$, ${\bf x}$ from the equation of motion (\ref{EOM}), 
we obtain the equation which is different from (\ref{neglect1}): 
\begin{align}
\mu_0\mathcal{H}^2\frac{\partial}{\partial T}\varphi(x)=\big(a^3\int^\infty_tdt'\ a'^{-3}\big)^{-1}\frac{\partial}{\partial t}\varphi(x)=-\frac{\partial}{\partial\varphi}V(\varphi(x)). 
\label{neglect2}\end{align}
Note that the time independence of $\mu_0$ is crucial in the above discussion while the value of $\mu_0$ itself does not contribute to the equation (\ref{neglect2}). 

Identifying the contribution from sub-horizon scales $\varphi_\text{UV}$ as the source term of the dynamics at super-horizon scales $\bar{\varphi}$: 
\begin{align}
\big(a^3\int^\infty_tdt'\ a'^{-3}\big)^{-1}\frac{\partial}{\partial t}\big\{\bar{\varphi}(x)+\varphi_\text{UV}(x)\big\}=-\frac{\partial}{\partial\bar{\varphi}}V(\bar{\varphi}(x)), 
\end{align}
\begin{align}
\varphi_\text{UV}(x)\equiv &\int \frac{d^3p}{(2\pi)^3}\ \theta(p-Ha)
\big[-i\frac{2^{\nu_*-1}\Gamma(\nu_*)}{\sqrt{\pi}}\frac{\big\{(1-\epsilon_*)H_*a^{\epsilon_*}\big\}^\frac{1}{1-\epsilon_*}}{p^{\nu_*}}e^{+i{\bf p}\cdot{\bf x}}a_{\bf p}+\text{(h.c.)}\big], 
\label{UV2}\end{align}
we obtain the Langevin equation which agrees with the result derived by the resummation formula (\ref{langevin2})-(\ref{noise2}): 
\begin{align}
\dot{\varphi}(x)=\dot{\bar{\varphi}}_0(x)-\big(a^3\int^\infty_{t}dt'\ {a'}^{-3}\big)\frac{\partial}{\partial\varphi}V(\varphi(x)). 
\label{same2}\end{align}
Here we made use of the identity between (\ref{linear2}) and (\ref{UV2}): $\dot{\varphi}_\text{UV}=-\dot{\bar{\varphi}}_0$. 

\section{Conclusion}\label{Conclusion}
\setcounter{equation}{0}

In accelerating universes, sub-horizon modes with given comoving momenta go outside of the horizon with cosmic evolution and then the degrees of freedom at super-horizon scales accumulate. 
As a consequence of the sensitivity to the accumulation, the propagator of a massless and minimally coupled scalar field acquires a growing time dependence. 
In interacting field theories with the scalar field, some physical quantities acquire secular growths which spoil perturbative investigations eventually. 

These IR effects are nonlocal in the sense that they are expressed by time integrals of functions of $H$, $\epsilon$. 
Therefore, they cannot be subtracted by local counter terms. 
Especially when the slow-roll parameter is time independent, the nonlocal effects can be identified as breakdown of the scaling law.

In order to evaluate the IR effects nonperturbatively, we extended the resummation formula of the leading IR effects in a general accelerating universe. 
Substituting the leading IR behaviors of the free field $\varphi_0$ and the retarded propagator $G^R$, 
the Yang-Feldman formalism is reduced to the Langevin equation with the white noise term (\ref{langevin2})-(\ref{noise2}).   
The difference from the Langevin equation in dS space is expressed as the slow-roll parameter dependence of the coefficient of each term. 

The white noise term is a direct consequence of the fact that quantum fluctuations of a massless and minimally coupled scalar field are frozen at super-horizon scales. 
The nonlinear term is expressed as the local operator $\partial V/\partial\varphi(\varphi(x))$. 
That is because up to the leading IR effects, the retarded propagator is spatially local and the time dependence of the vertex integral appears only in its upper bound. 

The Langevin equation can be translated to the Fokker-Planck equation (\ref{FP2})-(\ref{A2}). 
The final state solution of the Fokker-Planck equation is necessary to evaluate eventual contributions to physical quantities. 
T. Prokopec found that if the slow-roll parameter is constant and the interaction potential contains only dimensionless couplings, we can obtain an analytical solution of the final state \cite{Prokopec2015-2}. 
In a general case, an analytical solution may not always exist. 
In the situation that the time variations of $H$, $\epsilon$ are negligible during the IR effects grow, we derived the approximate solution of the final state (\ref{approximate}).  

Furthermore we derived the same Langevin equation by the semiclassical description of the scalar field. 
In the semiclassical approach, we identify the contribution from sub-horizon scales as the noise term. 
As seen in (\ref{linear2}) and (\ref{UV2}), we have only to flip the sign of the step function variable to obtain the same noise term with the resummation formula.  
The procedure for the noise term is the same as adopted in \cite{Starobinsky1986,Starobinsky1994,Starobinsky2009,Starobinsky2010-1,Starobinsky2010-2,Starobinsky2015}. 

In contrast, we adopt a different procedure in eliminating the second derivative terms from the equation of motion. 
If we neglect the second derivatives with respect to $t$, ${\bf x}$ in the same way as \cite{Starobinsky1986,Starobinsky1994,Starobinsky2009,Starobinsky2010-1,Starobinsky2010-2,Starobinsky2015}, 
the nonlinear term of the Langevin equation has a different coefficient with the resummation formula. 
To resolve the inconsistency, we choose the time coordinate $T$ as its friction term is constant in (\ref{friction}). 
In terms of $T$, we can compare $\frac{\partial}{\partial T}\varphi$ directly with $\frac{\partial^2}{\partial T^2}\varphi$.  
Neglecting the second derivatives with respect to $T$, ${\bf x}$, the nonlinear term of the Langevin equation has the same coefficient with the resummation formula. 

From these facts, we can conclude that as far as we adopt the appropriate time coordinate $T$, 
the semiclassical description of the scalar field is equivalent to the resummation of the leading IR effects, 
not only in dS space but also in a general accelerating universe.   

\section*{Acknowledgment}

This work was supported by Grant-in-Aid for Scientific Research (B) No. 26287044. 
H. K. would like to thank International Institute for Physics, which organized the international workshop `Quantum Fields and IR Issues in de Sitter Space'. 
The discussion during the workshop improved the quality of this paper. 


\end{document}